**Optimizing the $U_{eff}$ value for DFT+U calculation of paramagnetic solid-state NMR shifts by double Fermi-contact-shift verification**

Y. Liu, L. Zeng, C. Xu, F. Geng, M. Shen, Q. Yuan, B. Hu*

Shanghai Key Laboratory of Magnetic Resonance, State Key Laboratory of Precision Spectroscopy, School of Physics and Materials Science, East China Normal University, Shanghai 200062, P.R. China

E-mail: bwhu@phy.ecnu.edu.cn

Abstract: The isotropic chemical shifts can be calculated either by full-electron configuration, or by hybrid functionals, which costs a large amount of computational resources. To save the time, DFT+U could be employed to calculate the isotropic chemical shifts. However, the calculated properties are very sensitive to the Hubbard correction value $U_{eff}$. Here the double Fermi-contact-shift verification approach with DFT+U method is proposed with much higher computational efficiency, that is, simultaneously calculate the Fermi-contact shifts on two nuclei ($^6$Li and $^{17}$O) to predict the optimal $U_{eff}$. The optimal $U_{eff}$ is also helpful to the calculations of quadrupolar coupling constant $C_Q$, $g$-factor, band structure and density of states.

## 1. Introduction

In recent years, solid-state nuclear magnetic resonance spectroscopy (SS-NMR) has become a vital analytical method that provides atomic-level structural information of materials used for Li-ion batteries.[1] However, electrode materials suitable for battery applications always show a certain degree of disorder, for example caused by defects or by stochastic occupation of certain crystallographic sites, which makes the interpretation of NMR spectra to be challenging.[2] Especially, the cathode material of the Li-ion battery usually has paramagnetic sites with unpaired electrons, leading to a hyperfine effect and thus further increases the difficulty of analyzing the NMR shift. Fortunately, theoretical simulations of chemical shielding can provide good support and facilitate the interpretation of experimental NMR spectra.[3] Meanwhile, such theoretical calculations also pose a considerable challenge because of the large hyperfine shifts and the paramagnetic line broadening.[4] The isotropic hyperfine coupling interaction can be decomposed into Fermi-contact (FC) and pseudo-contact (PC, see Eq. 1) interactions.

DFT calculations with different functionals have been performed to obtain chemical shielding tensor and unambiguously assign experimental NMR signals.[5-9] For example, the isotropic chemical shift which contains several contributors was calculated by either hybrid functionals mixing different kinds of LDA/GGA functionals with Hartree-Fock (HF) exchange or full-electron configuration which uses full-potential linearized augmented plane-wave (LAPW) plus local orbitals.[10, 11] These methods are generally accurate enough, but may demand numerous computation resources.

In contrast to LAPW and hybrid DFT approaches, the DFT+U method is much more economical and provides good accuracy. Although the DFT functional (LDA/GGA) tends to poorly describe systems with localized transition-metal $d$ (or $f$) orbitals, such deficiency could be usually corrected by the DFT+U method that characterizes better strong intra-atomic interaction.[12] DFT+U has been widely accepted for studies of the nuclear magnetic properties of transition-metal oxides,[13, 14] and such an approach was also used for the calculation of chemical shift to qualitatively assign the NMR signals in several other types of cathode materials.[15, 16] Currently, a big challenge in the application of DFT+U for assigning the NMR signals is the determination of $U_{eff}$ values since the calculated Fermi-contact shifts and then the isotropic chemical shift are very sensitive to the $U_{eff}$ values.

In previous research, the Hubbard correction $U_{eff}$ was determined by several methods. One of the earliest works is the constrained local density approximation (cLDA) approach[17-21] where the $U_{eff}$ is calculated from the total energy variation with respect to the occupation number of the localized orbitals. A different approach based on the random-phase approximation (RPA) was later introduced[22], which allows for the calculations of the matrix elements of the $U_{eff}$ and its energy dependence. A commonly utilized approach is to verify the calculated bandgaps against the results of experiments or high-accuracy DFT calculations. Researchers usually test a variety of $U_{eff}$ values and choose the one which leads to the best bandgap outcome which is consistent with that from the experiment or hybrid DFT or GW calculations. By adopting this method, errors could come from inaccuracy of experiments and insensitivity in bandgap with a changing $U_{eff}$ value.

In this work, we focus on the development of a simple and fast scheme that permits the prediction of Hubbard correction $U_{eff}$ by calculating the large hyperfine interaction in paramagnetic SS-NMR. The $^6$Li and $^{17}$O NMR shifts of the model sample Li$_2$MnO$_3$ are simultaneously calculated using DFT+U with different $U_{eff}$ values, from which we obtained the calculated Fermi-contact shifts. Moreover, we systematically examined the effect of Hubbard $U_{eff}$ correction on the predicted hyperfine interaction of both $^6$Li and $^{17}$O, and then chose the optimal $U_{eff}$ which leads to linear relationship between the experimental Fermi-contact shifts and the calculated ones. This new approach to optimize $U_{eff}$ is much faster and should be applicable to other transition metal layered oxides.

## 2. Computation Methods

First-principle calculations were carried out on basis of the density functional theory (DFT) with the Vienna Ab initio Simulation Package (VASP) [23]. Spin-polarized generalized gradient approximation (GGA) calculations were performed with the Perdew-Burke-Ernzerhof (PBE)[24] exchange-correlation functional. The Li(1$s$,2$s$,2$p$), Mn(3$d$,4$s$), and O(2$s$,2$p$) orbitals are treated as valence states. To characterize the electronic and geometric structures of Li$_2$MnO$_3$, DFT+U method with different $U_{eff}$ values (ranging from 0 eV to 6 eV) was chosen for Mn atom to introduce a mean-field Hubbard-like term to improve the description of the electronic correlations for localized

*d* electrons. Scalar-relativistic norm-conserving pseudopotentials with nonlinear core correction were used, while the all-electron information was reconstructed using projector augmented wave (PAW)[25]. An energy cut-off of 600 eV was imposed for the plane wave basis. The Brillouin zone was sampled using a gamma centered grid with a k-point spacing finer than $2\pi \times 0.03$ Å$^{-1}$ for all calculations.

The structures were optimized via full relaxation of lattice parameters as well as atomic positions, until the residual forces were smaller than 0.02 eV/Å and the energy criteria was less than $10^{-8}$ eV. The gauge-including projector augmented wave (GIPAW)[26] approach within VASP code was used for the calculation of the NMR chemical shifts and the hyperfine tensors with a higher converging criterion of $10^{-9}$ eV was used for electronic minimization.

The EPR g-factors for the same systems were calculated by the Quantum-ESPRESSO (QE)[27] with PBE+U exchange functional. Scalar-relativistic norm-conserving pseudopotentials with nonlinear core correction were used, and then the all-electron information was reconstructed with PAW and the gauge-including projector augmented wave (GIPAW) [28]. The same energy tolerance and k-mesh sampling as mentioned above were used, while the plane wave cutoff energy was increased to 1600 eV for better accuracy.

## 3. Results and discussion

To correct the self-interaction error in the GGA formalism, a Hubbard $U_{eff}$ parameter was included for the Mn ions to treat the 3*d* correlations using the approach proposed by S. L. Dudarev[29], where the Coulomb matrix (*U*) and the exchange matrix (*J*) are combined to give an overall effective value $U_{eff} = U - J$. In this work, the value of *J* was set to 0. Here the calculation of hyperfine coupling interaction that dominates the chemical shifts was employed for determining the $U_{eff}$.

Li$_2$MnO$_3$ (space group: *C2/m*) was chosen as a model system for which the experiment $^6$Li, $^7$Li, and $^{17}$O NMR spectra are available for examining the accuracy of DFT calculated NMR shifts. Due to higher resolution of $^6$Li NMR as compared to that of $^7$Li, the calculated $^6$Li chemical shifts are reported here. For Li$_2$MnO$_3$, the Weiss temperature is -34 K, [30] the chemical shift for $^6$Li are 1461, 755, 734 ppm at 323 K[31] and the $^{17}$O chemical shift are 2264.7 and 1883.8 ppm at 328K[32], which is close to previously reported value[33]. For the reference sample of Li$_2$CO$_3$, the experimental $^6$Li chemical shift is 0 ppm and $^{17}$O chemical shift is 163.35 ppm, and the calculated $^6$Li chemical shift is -82.33 ppm and $^{17}$O chemical shift is 198.19 ppm. Here the $^{17}$O chemical shift of Li$_2$CO$_3$ was set to the average chemical shift of the two oxygen sites.

For the calculation, we first optimized the lattice parameters of Li$_2$MnO$_3$ using the PBE+U method with different $U_{eff}$ values, and then calculated the NMR chemical shift, bands, and density of states (DOS) if needed based on the optimized cell.

The chemical shift is calculated by[34]

$$\delta_{iso}^{cal} = \delta_{orbit}^{cal} + f \cdot \left(g_e \cdot A_{FC} + \Delta g_{iso} \cdot A_{FC} + \frac{1}{3} Tr[\Delta g_{aniso} \cdot A_{Dip}]\right) + \delta_{QIS}^{cal} \qquad (1)$$

where $f = \frac{S(S+1)\beta_e}{3k_B g_N \beta_N} \cdot \frac{1}{T-\theta}$, $S$ is effective electron spin, $\beta_e$ and $\beta_N$ are Bohr magneton and nuclear magneton respectively, $k_B$ is Boltzmann's constant, $T$ is the absolute temperature (K), $\theta$ is the Curie-Weiss temperature, $g_e$ and $g_N$ are the free-electron and nuclear g-factors, respectively, $A_{FC}$ and $A_{Dip}$ are the isotropic Fermi-contact (FC) and anisotropic traceless spin-dipolar contribution to the A-tensor (hyperfine coupling tensor) respectively, while $\Delta g_{iso}$ and $\Delta g_{aniso}$ are the isotropic and anisotropic terms of g-factor ($g = g_e + \Delta g_{iso} \cdot 1 + \Delta g_{aniso}$). $\frac{1}{3} Tr[\Delta g_{aniso} \cdot A_{Dip}]$ is commonly referred to as the pseudo-contact shift (PC), which is derived from the coupling between the nonrelativistic dipolar component of the hyperfine tensor and the $g$ anisotropy due to spin-orbit coupling. In our calculations, the $\delta_{orbit}^{cal}$ term of the sample was calibrated in relative to the reference sample Li$_2$CO$_3$ according to $\delta_{orbit}^{cal} = \sigma_{orbit}^{sample-cal} - \sigma_{orbit}^{ref-cal}$. The $\delta_{QIS}^{cal}$ term is the quadrupolar induced shift caused by the electric field gradient (EFG). For $^6$Li nucleus, $\delta_{QIS}^{cal}$ is very small, therefore this term could be omitted. However, for $^{17}$O nucleus, $\delta_{QIS}^{cal}$ should be included[35] and the values are ca. 10-20 ppm.

Therefore, Eq. (1) can be arranged as

$$\delta_{iso}^{cal} = \left(\sigma_{orbit}^{sample-cal} - \sigma_{orbit}^{ref-cal}\right) + f\left(g_e \cdot A_{FC} + \Delta g_{iso} \cdot A_{FC} + \frac{1}{3}Tr[\Delta g_{aniso} \cdot A_{Dip}]\right) + \delta_{QIS}^{cal} \quad (2)$$

which indicates that

$$f(g_e + \Delta g_{iso}) \cdot A_{FC} = \delta_{iso}^{cal} - \left(\sigma_{orbit}^{sample-cal} - \sigma_{orbit}^{ref-cal}\right) - \frac{1}{3}fTr[\Delta g_{aniso} \cdot A_{Dip}] - \delta_{QIS}^{cal} \quad (3)$$

For convenience, $f(g_e + \Delta g_{iso}) \cdot A_{FC}$ and $\frac{1}{3}fTr[\Delta g_{aniso} \cdot A_{Dip}]$ would be referred to as $\delta_{FC}^{cal}$ and $\delta_{PC}^{cal}$ respectively, and then we have

$$\delta_{FC}^{cal} = \delta_{iso}^{cal} - \delta_{orbit}^{cal} - \delta_{PC}^{cal} - \delta_{QIS}^{cal} \quad (4)$$

We tried to compare the calculated value $\delta_{FC}^{cal}$ with the experimental one $\delta_{FC}^{expt}$, however it is difficult to obtain the experimental values $\delta_{orbit}^{expt}$ and $\delta_{PC}^{expt}$. We approximate $\delta_{orbit}^{expt}$ by $\delta_{orbit}^{cal}$, $\delta_{PC}^{expt}$ by $\delta_{PC}^{cal}$ and $\delta_{QIS}^{expt}$ by $\delta_{QIS}^{cal}$, which makes

$$\delta_{FC}^{expt} = \delta_{iso}^{expt} - \delta_{orbit}^{expt} - \delta_{PC}^{expt} - \delta_{QIS}^{expt} \approx \delta_{iso}^{expt} - \delta_{orbit}^{cal} - \delta_{PC}^{cal} - \delta_{QIS}^{cal} \quad (5)$$

due to the fact that the calculated value $\delta_{orbit}^{cal}$ by GIPAW is accurate enough and much close to the $\delta_{orbit}^{expt}$, together with the fact that $\delta_{PC}^{cal}$ is usually ca. 1-5 ppm, which is much smaller than the value of $^6$Li $\delta_{iso}^{expt}$ and $^{17}$O $\delta_{iso}^{expt}$. Furthermore, it should be mentioned that $\delta_{QIS}^{cal}$ for $^6$Li is close to zero and neglectable, and that for $^{17}$O is around ca. 20 ppm which is very small compared with $\delta_{FC}^{expt} \sim 2000$ ppm.

The optimized structure of Li$_2$MnO$_3$ is shown in Fig. 1a, it can be seen that there are three Li atom sites and two O atom sites in the unit cell. Therefore, three chemical shifts of Li atom and two chemical shifts of O atom are identified in our calculations. Fig. 1b shows the calculated Fermi-contact shifts $\delta_{FC}^{cal}$ on both $^6$Li and $^{17}$O versus to the $U_{eff}$ value. The calculated chemical shifts change significantly from ca. 4838 ppm to –2118 ppm when $U_{eff}$ increases from 0 eV to 6 eV. In contrast, $^6$Li NMR doesn't undergo such dramatic changes in absolute values. It is obviously that inappropriate $U_{eff}$ may not give reliable aid on NMR signal assignment. Such an observation is not surprising, because the oxygen atoms are in close proximity to the Mn ions with $p$ orbitals pointing towards the occupied *d* orbitals, leading to much larger transferred spin density around the oxygen nucleus.

It is expected that an accurate $U_{eff}$ should give rise to a linear relation between the calculated $\delta_{FC}^{cal}$ and experimental $\delta_{FC}^{expt}$ of $^6$Li and $^{17}$O at the same time. From Fig. 1b, we can see that the $\delta_{FC}^{cal}$ and $\delta_{FC}^{expt}$ of $^6$Li and $^{17}$O show good linear relationship only when the $U_{eff}$ values are between 2.0 and 3.0, which was guided by a dash line in Fig. 1b. To find a more accurate $U_{eff}$ value, we further investigated the evolution of $\delta_{FC}^{cal}$ with $U_{eff}$ between 2.3 and 2.9 with a much smaller stepsize of 0.1. (Fig. 2) On the basis of the linear fitting, we found that $U_{eff}$ = 2.6 eV leads to the best linear correlation (R = 0.9996) between $\delta_{FC}^{expt}$ and $\delta_{FC}^{cal}$. It should be noted that the determination of the optimal $U_{eff}$ value is highly dependent on the properties of the system. For example, WANG et al.[36] studied the effect of $U_{eff}$ on the calculated oxidation entropy and found that $U_{eff}$ of 3.5 eV showed the best agreement with the experimentally observed oxidation state of Mn (Mn$^{4+}$). However, $U_{eff}$ = 5 eV was employed in other works to study the properties of lithium intercalation voltages and doping effects of Li$_2$MnO$_3$.[37,38] Here we emphasize that the optimal $U_{eff}$ value deduced in our work may not be applicable for reproducing properties other than NMR. However, we believe that this approach could be applied for various other materials.

The quadrupole coupling constant $C_Q$ is another chemical analysis technique related to NMR. In the next, we studied the effect of $U_{eff}$ on the calculation of $C_Q$ of $^{17}$O in Li$_2$MnO$_3$. The calculated $C_Q$ values of the two types of $^{17}$O in the unit cell of Li$_2$MnO$_3$ are shown in Fig. 3a. We can see the $C_Q$ values increase almost linearly with the $U_{eff}$. The calculated larger $C_Q$ of $^{17}$O is 4.4 MHz at the $U_{eff}$ value of 2.6 eV, which is consistent with previous reported value (4.6 MHz).[33] This means the optimal $U_{eff}$ value for chemical shift calculation is also good for $C_Q$ calculation. Moreover, the $g_{iso}^{cal}$ is also calculated by using different $U_{eff}$, and it also increases with the increase of $U_{eff}$ (Fig. 3b). When $U_{eff}$ = 2.6 eV, the value of $g_{iso}^{cal} = 1.926$ is obtained, which is close to the experimental value of $g_{iso}^{expt} = 1.994$,[39] suggesting the $U_{eff}$ = 2.6 eV could be a reasonable choice to accurately describe the $g_{\text{iso}}$ values and the isotropic chemical shifts. It should be noted that the calculated $g_{iso}^{cal}$ is roughly the same from $U_{eff}$ = 2.6 eV to 5 eV.

To show the effect of $U_{eff}$ on electronic properties of Li$_2$MnO$_3$, we further examined the band structure and DOS of Li$_2$MnO$_3$ using DFT+U with $U_{eff}$ of 2.6, 3.5 and 5.0 eV, respectively. Fig. 4 presents the band structure of Li$_2$MnO$_3$ calculated with increasing $U_{eff}$ value. The calculated band structures are almost the same, with the conduction-band minimum located at M-point and the valence-band maximum located at Y-points. The calculated band gap of Li$_2$MnO$_3$ using $U_{eff}$ of 2.6, 3.5 and 5.0 eV are 1.96, 1.99 and 1.66 eV, respectively, which are close to the experimental band gap of ~2-3 eV [40]. In addition, the partial density of states (PDOS) in Fig. 5 shows significant contributions of the O 2$p$ orbitals to the vicinity of the Fermi level. It also shows that the contribution of Mn 3$d$ orbitals are very sensitive to the $U_{eff}$, which coincides well with our calculated Fermi-contact shifts that is governed by the delocalization of unoccupied electrons. Overall, the $U_{eff}$ =2.6 gives reasonable band structure and PDOS of the Li$_2$MnO$_3$.

## 4. Conclusion

We predict the optimal $U_{eff}$ values, by simultaneously modeling the Fermi-contact shifts on different nucleus ($^6$Li and $^{17}$O) with respect to different $U_{eff}$ values, with Li$_2$MnO$_3$ as a model system. This double Fermi-contact-shift verification approach for DFT+U method allows well predicting the optimal Hubbard correction value $U_{eff}$. Our calculation shows that $U_{eff}$ predicted by NMR shifts is close to the reported $U_{eff}$ values in the literature. The use of accurate $U_{eff}$ is crucial for reproducing the experimental paramagnetic SS-NMR parameters (isotropic chemical shifts and the quadrupolar coupling constants $C_Q$) and the $g_{\text{iso}}$ values. $U_{eff}$ is also important to predict the band structure and density of states. This new approach to predict $U_{eff}$ can be applied to the DFT+U calculation for paramagnetic solids, together with other emerging materials.

**ACKNOWLEDGMENTS**

This work is supported by the National Natural Science Foundation of China (21522303, 21373086, 21703068). Y.L and L.Z are grateful for financial support for undergraduate students. Authors also thank ECNU Public Platform for Innovation (001) at East China Normal University for providing computing resources and acknowledge the Synchrotron XRD experiment supported by Shanghai Synchrotron Radiation Facility (BL14B1).

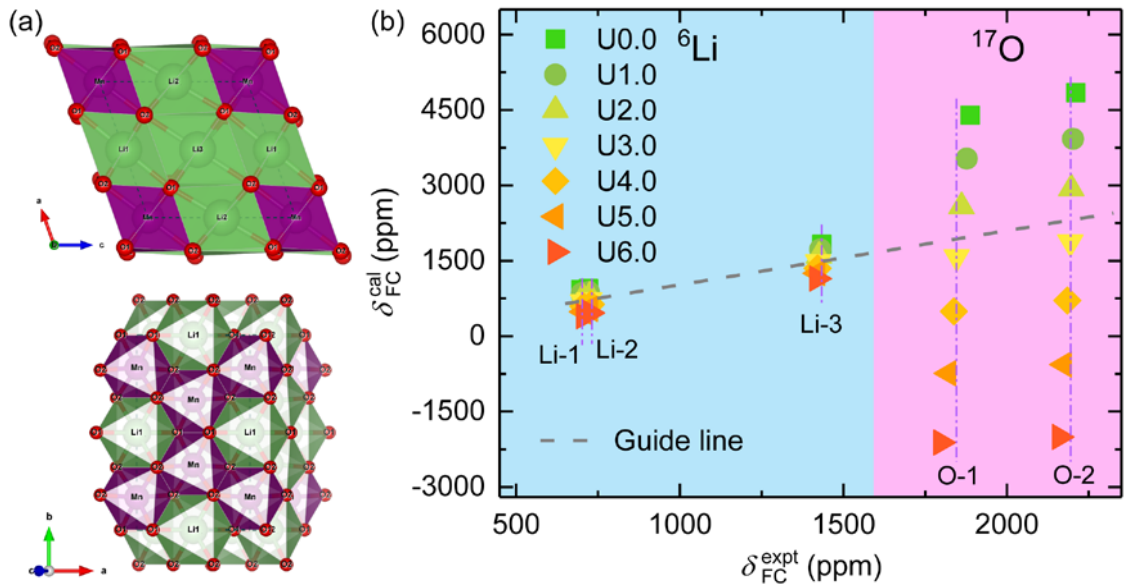

Figure 1(a). Schematic representations of the unit cell of $Li_2MnO_3$ (space group C2/m) with three Li atom sites and two O atom sites. The 4h, 2c, 2b and 4i, 8j sites are corresponding to Li-1, Li-2, Li-3 and O-1, O-2. (b). Calculated versus experimental Fermi-contact shifts of $Li_2MnO_3$. The Fermi-contact shifts of $^6Li$ and $^{17}O$ are plotted on the same figure with two different regions. $U_{eff}$ was applied with the value varied from 0.0 to 6.0 eV. A dash line is plotted to guide the eyes.

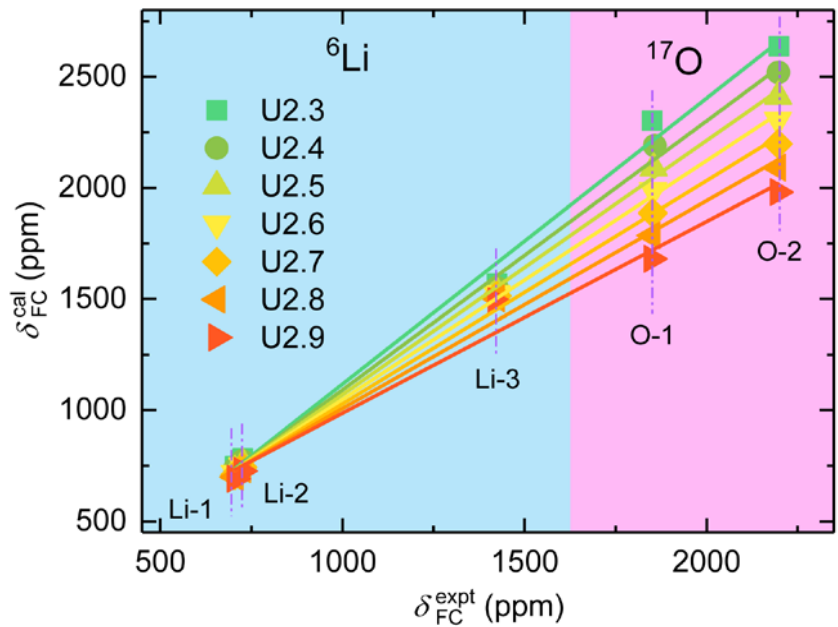

Figure 2. Calculated versus experimental Fermi-contact shift of $Li_2MnO_3$. The Fermi-contact shifts of $^6Li$ and $^{17}O$ are plotted on the same figure within two different regions. The $U_{eff}$ correction is indicated on the legend. Fitting R values for $U_{eff}$ = 2.3, 2.4, 2.5, 2.6, 2.7, 2.8, 2.9eV are 0.9963, 0.9981, 0.9992, 0.9996, 0.9991, 0.9973, 0.9934, 0.9865.

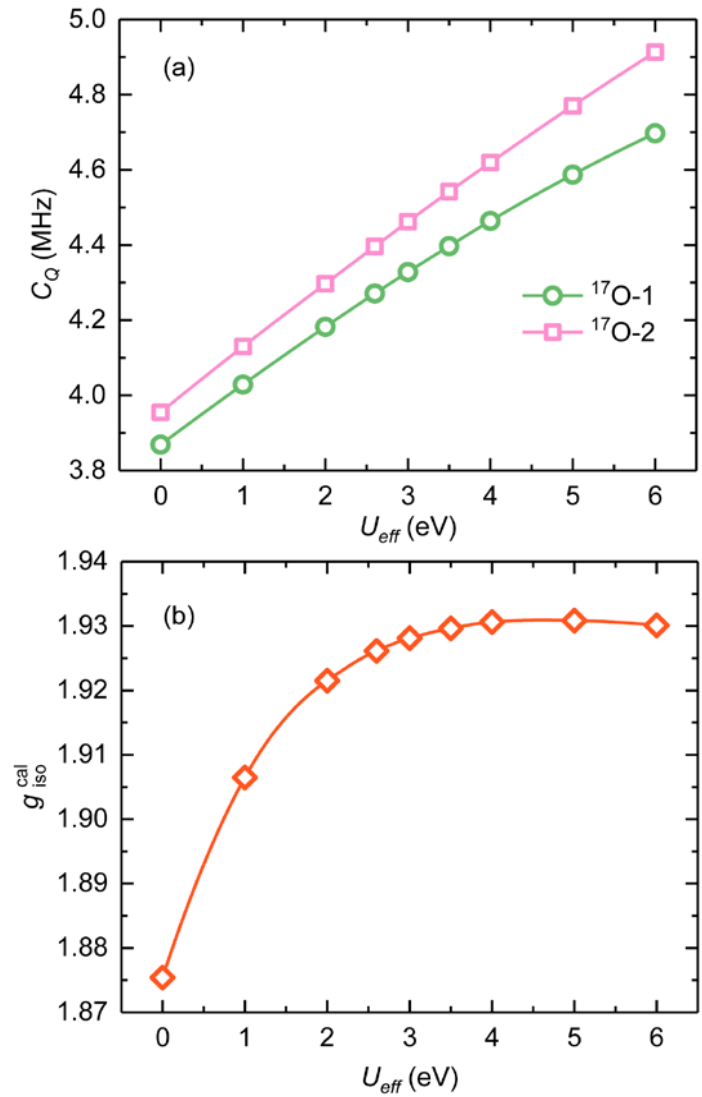

Figure 3. Calculated $C_Q$ of $^{17}O$ (a) and calculated $g_{iso}^{cal}$ of $Li_2MnO_3$ (b) using PBE+U with $U_{eff}$ varying from 0.0 to 6.0 eV.

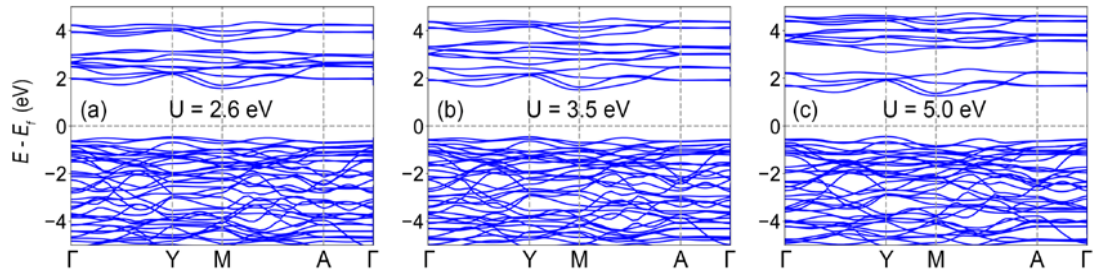

Figure 4. Calculated band structure of Li$_2$MnO$_3$ for GGA+U with $U_{eff}$ = 2.6(a), 3.5(b) and 5.0 (c) eV. All energies refer to the Fermi energy. Bandgaps are 1.96, 1.99 and 1.66 eV respectively from left to right.

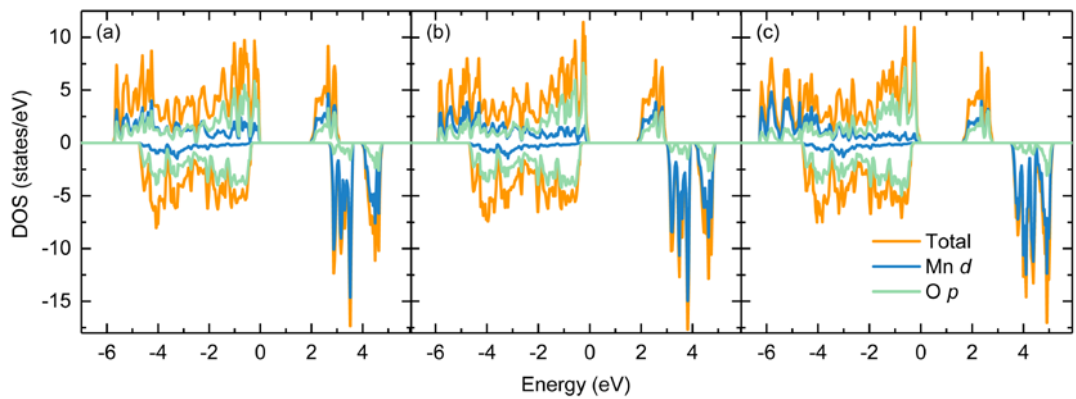

Figure 5. Calculated partial density of states (DOS) of Li$_2$MnO$_3$ with GGA+U ($U_{eff}$ = 2.6(a), 3.5(b), 5.0(c) eV, from left to right). The contributions from Mn $d$ orbitals and O $p$ orbitals are labelled in blue or green.